\documentclass{aa}
\usepackage[dvips]{graphicx}
\def\simless{\mathbin{\lower 3pt\hbox
{$\rlap{\raise 5pt\hbox{$\char'074$}}\mathchar"7218$}}}   
\def\simmore{\mathbin{\lower 3pt\hbox
{$\rlap{\raise 5pt\hbox{$\char'076$}}\mathchar"7218$}}}   
\newcommand{\be}{\begin{equation}}
\newcommand{\ee}{\end{equation}}

\begin{document}
\title{Excitation of low-frequency QPOs in black hole accretion}
\titlerunning{Excitation of low-frequency QPOs}
\author{Dimitrios Giannios \inst{1,2,3}
\and Henk C. Spruit \inst{1}
}

\institute{Max Planck Institute for Astrophysics, Box 1317, D-85741 Garching, Germany
\and University of Crete, Physics Department, P.O. Box 2208, 710 03, 
Heraklion, Crete, Greece
\and Foundation for Research and Technology-Hellas, 711 10, Heraklion, Crete, Greece
}

\offprints{dgiannios@physics.uoc.gr}
\date{Received / Accepted}

\abstract{
We study possible mechanisms of excitation of quasiperiodic oscillations in the accretion 
flow of black hole accreters in their hard spectral states, in the context of the `truncated 
disk' model. Quasi-spherical oscillations of the inner ion-supported accretion flow (ISAF) can 
be excited by the interaction of this hot flow with the cool disk extending outward from it. 
The fundamental mode of (p-mode) oscillation is most easily excited, and has a frequency near 
the Kepler frequency at the inner edge of the cool disk. The strongest excitation mechanism is 
a feedback loop involving cooling of the ISAF by soft photons from the cool disk and heating of 
the cool disk by the ISAF, while synchrotron emission can be a relatively strong damping effect. 
Growth times are computed by detailed Comptonization calculations of the interaction of an idealized 
ISAF with a cool disk. Typical growth times as short as a few dynamical times are found, while 
amplitudes can be as large as 10\%.

}

\maketitle

\section{Introduction} 
\label{intro}

A common feature of the X-ray light curves of black hole binaries (most of them transient X-ray
 sources) is quasiperiodic oscillations (QPOs)  (narrow power peaks with high quality factor 
$Q\simmore2$ (i.e. Belloni et al. 2002)). QPOs can be categorized -according to their peak 
frequencies- as high-frequency QPOs (HFQPOs) and low-frequency QPOs (LFQPOs). HFQPOs appear 
at $\nu\simmore 60$ Hz and have low rms amplitudes (typically 1\% of the mean count rate). 
They are typically observed when black hole binaries are in the very high state, i.e. when they 
exhibit both a thermal component at $k_BT\sim 1$ keV and a hard X-ray power law with photon 
energy index ($dN/dE\propto E^{-\Gamma}$) $\Gamma \sim2.5$. Because of their high frequencies, 
they are thought to convey information about physical properties that occur near the black 
hole horizon. A model which explains the 2:3 frequency ratio often seen in HFQPOs is given
in Rezzolla et al. (2003).

The more common  LFQPOs have a characteristic frequency range of 0.1-10 Hz,  
contribute significantly to the total variability (rms amplitudes of the order 10\% and higher
 have been seen in  XTE J1550--564 and XTE J1118+480) and have been observed in both the low/hard
and very high spectral states (for a review see Swank 2001). Their amplitude is generally higher 
at higher photon energies, though exceptions occur (for example in a QPO seen in  
XTE J1550--564 (Kalemci et al. 2001) and  in Cygnus X-1 (Rutledge et al. 1999)). 
Characteristic for the low frequency QPOs is a decrease of the frequency during the 
decay of the transient, suggesting a physical relation between frequency and accretion 
rate. 

 In recent years, it has become increasingly clear that the hard state power density spectrum (PDS)
of Cygnus X-1 and other black hole binaries shows several breaks, i.e. frequencies where the PDS
steepens (Nowak et al. 1999) and that this complex structure can be described with multiple
broad noise components (Nowak 2000; Nowak et al. 2002; Pottschmidt et al. 2003). Moreover, 
there is evidence of a correlation between the horizontal branch oscillations and lower kHz
QPOs in Z and atoll sources that can be extrapolated, by more than two orders of magnitude in
frequency, to include a correlation between LFQPOs and noise components in black hole binaries 
(Psaltis et al. 1999). In this work, we develop a model that attempts to clarify the origin of 
at least some of the LFQPOs. 

\subsection{QPOs as diagnostics of the accretion flow}
Apart from their intrinsic interest, the oscillations are a potentially important diagnostic of the still 
poorly understood inner regions of the accretion flow. In the `disk corona' model (e.g. Poutanen \& Fabian 
1999), a cool ($<1$ keV) disk extends to the innermost stable orbit around the hole, while the hard X-rays 
are produced in a magnetically heated corona above it.  In the `truncated disk' model, on the other hand, 
the cool disk extends inward only to a transition radius $R_{\rm tr}$, inside of which the accretion takes 
place as an ion supported flow (Rees et al. 1982). In such flows the ion temperature is near the virial 
temperature, while the electrons producing the hard X-rays have a temperature around $100$ keV (Shapiro et 
al. 1976; Ichimaru 1977). Their hydrodynamical properties have been studied by Gilham (1981) and in detail 
by Narayan and Yi (1994 and references therein) who called them ADAFs. To distinguish them from the radiatively
 supported accretion flows (also called ADAF), we use here the name `ion supported accretion flow' (ISAF).

The arguments for and against either of these geometries are part of a current debate in the literature. 
The main obstacle to acceptance of the truncated disk model is the difficulty in visualizing a process that 
can make a cool disk flow  to change into a hot ion supported flow, across a highly unstable region of 
intermediate 
temperatures. Without going into details, we note here that this difficulty has been overcome by Spruit \& 
Deufel (2002). It is shown there that viscous heating of the ions in the interaction region between the ISAF 
and the cool disk near the transition radius causes the cool disk to evaporate and feed the ISAF even when the 
transition radius is close to the last stable orbit.

Distinguishing the two accretion geometries observationally is not possible from the X-ray spectrum alone. 
The hope is that a prominent feature like the low-frequency QPO can help in settling the question of the 
accretion geometry. In the truncated disk model, the truncation radius provides a characteristic length 
scale and an (orbital) frequency scale that can be associated with the QPO. The dependence of frequency 
on accretion rate would then imply an increase of the transition radius with decreasing accretion rate. 
Supporting evidence for such a physical relation is found in the correlation between reflection indicators 
(Fe line strength and amplitude of the Compton reflection hump) and the strength of the soft component 
in the X-ray spectrum with QPO frequency (Zdziarski et al. 1999; Gilfanov et al. 2000).

\subsection{The importance of excitation}
The number of possible modes of oscillation in an accretion disk is as large as those of a star.
Discrete spectra of trapped oscillations can occur near the last stable orbit around a hole because
of the shape of the relativistic potential (Kato 1990; Honma et al. 1992; Ipser 1996; Nowak \&
Wagoner 1991, 1992). But even a disk in a Newtonian potential supports trapped oscillations near
its inner or outer edge (Drury 1980; Blaes \& Hawley 1988; Kaisig 1989). The spectra of such modes are 
quite complex (e.g.\ Narayan et al. 1987). It is not surprising, therefore, that picking modes to match 
an observation is not too difficult (e.g. Titarchuk et al. 1998).

To make progress in interpreting QPOs, criteria are needed to identify the most plausible modes. An 
important clue is the amplitude of the oscillation. To explain modulations in the total X-ray luminosity 
as high as 10\% with a hydrodynamic oscillation, the amplitude of the oscillation must be significant, 
not only at the larger distances where the frequency of the oscillation is presumably set, but also in 
the regions close to the hole where most of the X-rays are produced. Secondly, the observed flux is an 
average over the accretion flow, so that all nonaxisymmetric modes average out (except for strong 
nonlinearity, relativistic beaming, or optical depth effects). Similarly, the visibility of modes 
with one or more nodes in the radial part of their structure will be strongly suppressed. The most 
likely modes capable of producing large amplitudes in the X-ray flux are therefore axisymmetric modes 
with few nodes in the radial and vertical directions.

Finally, there must be a plausible mechanism to excite the mode. While broad band noise might be explainable 
as some consequence of turbulence in the flow (Lyubarskii 1997; Churazov et al. 2001), the high amplitudes 
seen in  LFQPOs probably require an excitation process intrinsic to the oscillation itself.

The wide and possibly continuous variation of the LFQPO with (inferred) accretion rate is natural 
in the truncated disk model. In addition, the presence of two components in the accretion flow around the 
transition radius gives rise to possible excitation mechanisms that are absent in disk+corona models.

\section{Excitation by interaction}

An ion supported accretion flow and a cool disk partially embedded in it interact through exchange of mass, 
energy and angular momentum. Exchange of mass takes place in the form of an evaporation process from some 
part of the cool disk, as needed to produce the ISAF in the first place, and possible condensation onto 
the disk at other locations. The evaporation can take place at a large distance where the virial temperature
 is of the order $10^6$--$10^8$ K, through processes similar to those maintaining the mass balance in the 
solar corona (Meyer et al. 2000; Meyer-Hofmeister \& Meyer 2001; Liu et al. 2002). The observations mentioned
 above, however, show that the cool disk is truncated at a much closer distance to the hole, at least in some
 cases. The mechanism that allows the disk to evaporate at such distances has been identified by Spruit \& 
Deufel (2002).

Thus assuming that the cool disk evaporates from a region near its inner edge, the resulting ISAF spreads 
both inward, feeding the hole and outward, spreading over the cool disk. At some distance from the inner 
edge, the spreading ISAF may condense back onto the cool disk, so that a region of coexistence between the 
cool disk and the ISAF is maintained.

An oscillation might be excited by the exchange of mass between the cool disk and the ISAF because 
the accretion time scale in the disk is much longer than the viscous spreading time of the ISAF.
Mass evaporating from the truncation radius $R_{\rm tr}$ depletes the surface density there. When this
mass condenses from the ISAF again on to the disk at some distance $R_{\rm out}>R_{\rm tr}$, it 
creates an enhancement there which takes a viscous time $(R_{\rm out}-R_{\rm tr})^2/\nu$ to arrive at
$R_{\rm tr}$. Spreading in the ISAF and condensation at $R_{\rm out}$ can then feed the next circle.

A second possibility for excitation is through interaction by angular momentum. A known example is viscously 
excited oscillation found by Kato (1978), and studied in detail by Honma et al. (1992); Papaloizou \& Stanley 
(1986) and others. In this mechanism, a sound wave oscillation is excited viscously in a rotating flow. The 
amplitude of the oscillation peaks at the outer edge of the flow (Kley et al. 1993). The boundary between 
the ISAF and the cool disk might be sufficient to excite a sound wave in the ISAF through this mechanism.

Finally, exchange of energy between the ISAF and the cool disk can excite oscillations. The flow is cooled 
by soft photons (either internally generated by synchrotron emission or externally from the cool disk) and 
heated by viscous dissipation. Changes in the soft photon input heat or cool the ISAF, causing it to expand 
and contract on a sound crossing time scale. Since the ISAF is partially pressure supported, the relative 
changes in size are of the same order as the relative changes in heating or cooling rate. The size of the 
region of overlap with the cool disk thus varies as well, and thereby the soft photon input into the ISAF.

In the following we investigate the conditions under which interaction by exchange of energy can produce 
self-excited oscillations. More specifically, we focus on the variation of radiative cooling of the flow, 
and assume that the viscous heating does not vary. The viscous heating can excite/damp oscillations on 
the viscous time scale. However, we show in section 5 that radiative cooling excites/damps oscillations 
on much shorter time scales, validating our assumption (for further discussion see  Sect. 5.3).

\section{Model}

We consider low order oscillation modes of the ISAF, excited by the energetic interaction
between the ISAF and the cool disk.  There are different interaction paths, and whether
these  excite or damp an oscillation of the ISAF depends on geometrical properties and on
the source of soft photons. 

\subsection{Geometry and soft photon sources}

\label{phsrc}
The basic ingredients of the model are a standard thin disk (Shakura \& Sunyaev 1973) truncated at a radius 
$R_{\rm tr}$ and an ISAF 
extending to a radius $R_{\rm out} > R_{\rm tr}$.
Inverse Compton scattering of soft photons is the natural radiative process to explain the non-thermal 
X-ray spectrum in the low/hard state, but there are several possible sources for the soft photons, 
depending on the geometry of the flow and model assumptions.

Soft photons can be generated internally in the ISAF as thermal synchrotron photons. This will be the 
dominant source of soft photons in an ion supported flow near a black hole if the magnetic field strength 
is some fraction of the equipartition field. This synchrotron radiation is strongly self absorbed below a 
characteristic turn-over frequency ${\nu}_{\rm t}$. Most of the synchrotron flux is emitted near this 
frequency, which typically lies in the infra-red or the optical region of the electromagnetic spectrum.

A second source of soft photons is the radiative interaction between the hot flow and the cool disk. 
Hard X-rays from the inner hot flow are partly absorbed and partly reflected by the cool disk. The 
absorbed energy is reprocessed into thermal soft flux and emitted by the disk. These soft photons 
are the seed photons Comptonized by the ISAF. The energetics and spectra produced in this 
{\em radiation-mediated} interaction  have been studied by Haardt \& Maraschi (1993).

In addition to this purely radiative interaction, the hot flow also heats the cool disk by `proton 
illumination' (Deufel et al. 2001; Deufel et al. 2002; Spruit \& Deufel 2002). In this process, part 
of the ISAF condenses onto the disk in the overlap region. Energetic protons (which carry most of the 
ISAF's energy) are slowed down by the disk plasma and their energy gets thermalized and appears as soft 
flux. The proton energy flux $F_{\rm p}$ which is released in the disk is of the order
\be
F_{\rm p}=n_{\rm p}(k_{\rm B}T_{\rm p})c_{\rm s},
\label{flux} 
\ee
where $c_{\rm s}$ is the proton sound speed and  $n_{\rm p}$ is the proton number density in
the ISAF. The density is of the order $n_{\rm p}=\tau/({\sigma}_T R_{\rm out})$, where $\tau
$ is the  Thomson optical depth of the ISAF  and ${\sigma}_T$ the Thomson cross
section.      

If half of this flux is re-emitted isotropically as soft photons, the disk surface temperature
$T_{\rm disk}$ is given by the relation $F_{\rm p}/2=\sigma T_{\rm disk}^4$, with
$\sigma$ the Stefan-Boltzmann constant ($\sigma=5.67\cdot10^{-5}$ erg cm$^{-2}$ s$^{-1}$ K$^{-4}$). We get
\be
T_{\rm disk}\simeq(\frac{\tau}{m_{1}r^{5/2}})^{1/4} \qquad \rm keV, 
\label{Tdisc}
\ee
where $m_{1}$ is the mass of the black hole in units of 10 solar masses ($2\cdot 10^{34} $g)
and $r$ the distance from the black hole in units of the  Schwarzschild radius  $R_{\rm s}=2GM/c^2$. 
We see that $T_{\rm disk}\sim100$ eV is a typical value for $\tau$ of order unity and $r\sim50$.
In the calculations reported below, the effects of each of these sources of soft photons on
the oscillation are investigated. 

\subsection{The oscillations, model approximations}

The excitation mechanisms considered here operate on the balance between heating and
cooling in the ISAF, in particular through a modulation of the soft photon input. The
simplest modes and the most likely ones to be excited by this mechanism are the
axisymmetric modes with few nodes in the radial direction, in particular the fundamental
mode.  In the following, we restrict ourselves to the fundamental mode of the ISAF. As in
a star, it has a frequency of the order of the Kepler  frequency at its outer edge, or the  inverse
time for a sound wave to travel  across it. 

Since the fundamental mode senses only the average sound speed, it is sufficient to model
the ISAF as spherical. Since the rotation period is of the order of the sound crossing time,
its effects on the fundamental mode are modest, hence we also ignore the effect of
rotation on the mode. Each of these approximations can be easily improved on with a
more detailed model of the structure of an ISAF. At the present level of understanding of
the geometry of the accretion flow onto black holes, this is not a primary concern.

We further simplify the model by taking the equilibrium electron temperature $T_{\rm e}$ and 
density $\rho_0$ as homogeneous. The fundamental mode of oscillation is approximated by a 
homologous expansion and contraction.

Thus, if $R(t,R_0)$ is the radial position of a shell of mass with equilibrium position $R_0$,
a sinusoidal oscillation of the model ISAF is described by
\be
R(t,R_0)=R_{0}[1+ a\cdot \sin {(\omega t)}],
\label{}
\ee
where $a$  is the fractional amplitude of the
oscillation. The radial velocity and the density perturbation are
\be
v(t,R_0)=a \omega R_0\cos (\omega t), \qquad \Delta \rho=-3 a{\rho}_0\sin(\omega t).
\label{wint}
\ee
Also the perturbation of the optical depth of the ISAF ($\tau =
\rho_{0} \kappa_{\rm es} R_{\rm out}$) is
\be
\Delta \tau = -2 a\tau \sin(\omega t).
\label{dtau}
\ee

Such a non-rotating, constant density configuration in homologous motion has a
fundamental frequency of oscillation (Cox 1980)
\be
\omega=\sqrt{(3\gamma -4)\frac{GM}{R_{\rm out}^3}},
\label{freq} 
\ee
where $\gamma$ is the ratio of specific heats, assumed constant. For the ion cloud of an ISAF, 
the non-relativistic value $\gamma=5/3$ is a good approximation, which will be used in this work. 
The oscillation frequency is then exactly equal to the Kepler frequency ${\omega}_k$. (In view 
of the approximations made, this exact agreement is a coincidence.)

\subsection{Excitation and damping}

\label{excit}

An efficient excitation mechanism is needed, since the observed oscillation is
often quite strong, accounting for a large fraction of rms variability.
In the present model, the excitation is due imbalance of cooling and heating terms
during the oscillation, such that the net work done on the gas by the pressure forces
is positive during an oscillation cycle.

With standard stellar oscillation theory (see for example  Cox 1980; Unno et al. 1979), the
rate of excitation of an oscillation can be expressed by the work integral 
\be   
\langle \dot W \rangle =-\frac{1}{T} \int_m dm \int_0^T
(\gamma-1)\frac{\Delta\rho}{\rho}\Delta \dot q dt.
\label{work}
\ee
 Where $\Delta \dot q$ is the (Lagrangian) variation of the net cooling rate per unit mass 
during the oscillation, the first integral is over the mass of the ISAF and the second over the
oscillatory period $T$. An ideal gas with constant ratio of specific heats $\gamma$ has
again been assumed.

If the excitation rate is small compared with the oscillation frequency (as needed in order
for the quality factor $Q$ of the oscillation to be high), the work function can be
represented by difference in the net cooling rate at the maxima and minima of the
oscillation (see Appendix A)

\be
\langle\dot W \rangle= 3(\gamma-1)a\frac{\dot q_{\rm exp}- \dot
q_{\rm contr}}{4},
\ee
where $a$ is the relative displacement amplitude as before, and $\dot
q_{\rm exp}$, $ \dot q_{\rm contr}$ are the net cooling rates at
maximum expansion and maximum contraction, respectively.

If cooling mechanisms are more efficient in the expanded phase of the
oscillation  than in the contracted phase, then in every oscillatory cycle there
is a net increase of the oscillation's energy proportional to
$\dot q_{\rm exp}-\dot q_{\rm contr} $ which means excitation of the
oscillation. If, on the other hand, $\dot q_{\rm exp}-\dot q_{\rm contr}<0$
then the oscillation gradually decays.

\subsection{Effect of different sources of soft photons}
To find out whether a mode can be excited by the processes that determine the energy
balance, special attention  must be paid to the cooling mechanisms of the ISAF, since these
are more easily modified than the viscous heating of the flow (cf.  Sect. 5.3).
Inverse Compton accounts for most of the cooling of the ISAF (at 
least for the conditions under which the observed spectra are
produced, since these are dominated by hard X-rays).
Soft flux, that crosses the hot and optically thin medium of the 
ISAF, is upscattered by energetic electrons resulting in
energy transfer from the electrons to the radiation field. 
The cooling rate of the ISAF is thus controlled by the input of
soft photons. As discussed above, we  consider three sources of 
soft photons: 

i) synchrotron photons from the inner parts of the flow,

ii) irradiation of the thin disk by X-rays from the ISAF, 

iii) proton illumination of the thin disk in the overlap area.

The strength of Compton cooling is determined by the  parameter,
$y=4 k_{\rm B}T_{\rm e}/(m_{\rm e}c^2)\tau$ 
and the soft flux: $ \dot{q_{\rm c}}=L_{\rm soft}(e^y-1)$.
During the expansion phase the optical depth of the ISAF drops  (Eq. [\ref{dtau}])
and so does the electron temperature (see next section), resulting in a
decrease of the $y$ parameter. The opposite trend is observed during
contraction. So, if the soft flux does not vary much, the cooling rate
of the expanded phase must always be less than the cooling rate of the
contracted phase. This leads to damping of the oscillations. On the
other hand if the soft flux increases quickly enough during expansion, 
there can be excitation of the oscillations.

When synchrotron photons are the `seed' photons to be Comptonized,
then it is straightforward to show that they act as a damping
mechanism. This is because the synchrotron flux increases strongly with 
electron temperature and the strength of the magnetic field and less
strongly with the Thompson optical depth (e.g. Wardzi\'nski \&
Zdziarski 2000). The
magnetic field scales as  $B^2\sim P_{\rm gas}$, which means
that $B$, $T_{\rm e}$ and $\tau$ all decrease during expansion. The
synchrotron soft flux decreases and so does the cooling rate during
expansion, resulting in energy loss of the oscillation. 

On the other hand, in the irradiation and proton illumination cases,
it is not a priori clear whether the net effect is damping or
exciting. Detailed calculations of the cooling rates are needed to
settle this question.                   
               
\section{Calculations}

\subsection{The electron temperature in equilibrium}

The energy source of the electrons in the two temperature plasma is 
the transfer of energy from the hot protons through Coulomb interactions.
This causes the protons to cool at a time scale  (Spitzer 1956; Stepney 1983) 
$t_{\rm p}$
\be
t_{\rm p}=\sqrt{\pi\over 2}\frac{m_{\rm p}}{m_{\rm e}}\frac{1}{n\sigma_{ T} c  
\ln \Lambda}\Big(\frac{k_{\rm B}T_{\rm e}}{m_{\rm e}c^2}\Big)^{3/2}  ,
\label{tp}
\ee
where $T_{\rm e}$ stands for the electron temperature, $n=\tau/(\sigma_{
\rm T} R_{\rm out})$ is the number density of
electrons, $\ln \Lambda$ is the Coulomb logarithm.  This expression
is the non-relativistic one and applies when $T_{\rm e}\ll T_{\rm p}\ll m_{\rm p}
T_{\rm e}/m_{\rm e}$. Where $T_{\rm p}$ is the proton temperature assumed to be
the virial temperature
\be 
T_{\rm p}=T_{\rm vir}=\frac{GMm_{\rm p}}{3k_{\rm B}R}\simeq
\frac{156}{r} \qquad \rm MeV.
\label{Tvir}
\ee
As will be shown in the rest of the section,
these inequalities hold for most of the parameter space of $T_{\rm e}$, $T_{\rm p}$
in which we are interested in this work.    

Because of their lower temperature, the heating time scale of the electrons is 
shorter than the proton cooling time scale
\be 
t_{\rm e}=\frac{T_{\rm e}}{T_{\rm p}}t_{\rm p}. 
\label{te}
\ee
The heating rate of the electrons (erg/sec) over the whole ISAF is then given by
$nk_{\rm B}T_{\rm e}V/t_{\rm e}$, with $V$ being the volume of the ISAF. In
equilibrium this quantity equals the cooling rate of the electrons,
which is assumed to be  controlled by Compton cooling and ultimately equal to the
observed X-ray flux of the source
\be
L_{\rm x}=\frac{nk_{\rm B}T_{\rm e}V}{t_{\rm e}}. 
\ee
Combining this equation with Eq. (\ref{te}), Eq. (\ref{tp}), $V=(4/3)\pi R_{\rm out}^3$ and taking 
$\ln \Lambda\simeq 20$, $T_{\rm p}=T_{\rm vir}$, we 
can relate the temperature to the optical depth and X-ray flux

\be
k_{\rm B}T_{\rm e}\simeq 5{\tau}^{4/3}\left({L_{\rm E}\over L_{\rm x}}\right)^{2/3}\quad
{\rm keV}
\label{telk}
\ee    
(Shapiro et al. 1976). Here $L_{\rm E}$ is the Eddington luminosity.

 A further constraint on $T_{\rm e}$ and $\tau$ from observable quantities comes
from the slope $s$ of the X-ray spectrum in a $\nu\cdot F_\nu$ diagram ($\nu\cdot F_\nu \sim
\nu^s$). With an analytic 
fit to Comptonized spectra given in Wardzi\'nski \& Zdziarski (2000) their Eq. [22], which 
agrees with reasonable accuracy with our computational results in  Sect. \ref{result}) we get
for  the hard X-ray power-law slope
\be
s\simeq 1-\frac{0.8}{{\tau}^{4/5}}{100{\rm keV}\over k_{\rm B}T_{\rm e}},
\\ 0.5\le\tau\le 2, \quad k_{\rm B}T_{\rm e}\le 200\,  {\rm keV}. \label{s}
\ee
Eqs. (\ref{telk}) and (\ref{s}) 
determine the electron temperature of the ISAF, expected from electron heating through
Coulomb interaction with virial protons, in terms of the optical depth $\tau$ of the flow
and two observables, the X-ray luminosity and the slope of the hard X-ray spectrum. 
This is shown in Fig.\ 1. 

 From Fig.\ 1 one can see that the electron temperature of the ISAF, if it is to account
for the observations, is $\sim 100$ keV. Furthermore since $T_{\rm p}=
T_{\rm vir}$, the inequalities $T_{\rm e}\ll T_{\rm p}\ll m_{\rm p} T_{\rm e}/m_{\rm e}$
 hold for $r_{\rm out}<1000$ validating the use of Eq.\ (\ref{tp}).

If the optical depth is expressed in terms of the accretion rate, with an assumed viscosity,
this can be further reduced to a dependence of the expected electron temperature on
accretion rate (e.g. Rees et al. 1982; Narayan \& Yi 1995; Zdziarski 1998). 

\begin{figure}
\resizebox{\hsize}{!}{\includegraphics[]{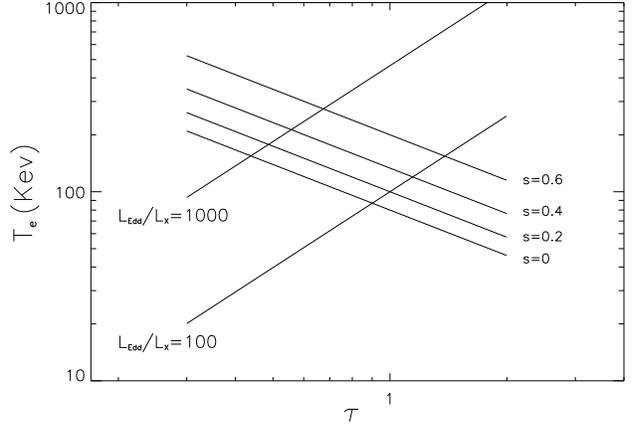}}
\caption[]{Electron temperature in an ion supported flow of optical depth $\tau$,
as a function of the observed X-ray luminosity $L_x$ and luminosity index  $s$ of the 
spectrum ($\nu F_{\nu}\sim \nu^{s}$).}  
\label{fig2}
\end{figure}

\subsection{The electron temperature during oscillation}

In order to calculate the excitation of an oscillation, the net cooling/heating of the flow has to be
followed during a cycle (Sect. 3.3). For this we need the variation of the electron temperature
during the oscillation. We compare the electron heating time scale with half of the oscillation
period $P$ (time between two passes through the equilibrium position). From equations
(\ref{tp}), (\ref{Tvir}), (\ref{te}) the heating time scale of the electrons is
\be
t_{\rm e}=6\cdot 10^{-7} \frac {m_{1} r_{\rm out}^2}{\tau} \Big(\frac{k_{\rm B}T_{\rm e}}{100 \rm 
keV}\Big)^{5/2}
\qquad \rm sec,
\label{} \\
\ee
  and 
\be
P/2=4.3\cdot 10^{-4} m_{1}r_{\rm out}^{3/2} \qquad \rm sec,
\label{}
\ee
where $r_{\rm out}$ is the radius of the ISAF in $R_{\rm s}$. It follows that 
$t_{\rm e}\ll P/2$ for optical depths $> 0.01$ and $r_{\rm out} \simless 1000$.

The electron temperature therefore adjusts quasistatically to the changing proton temperature, 
so equilibrium equations from the last section still apply during the oscillation. 
Taking the Lagrangian derivative ($\Delta$) of Eq. (\ref{te}) and  using 
$\Delta n=-3 a n\sin(\omega t)$ and $\Delta T_{\rm p}=-2T_{\rm p}\sin(\omega t)$
(see next section) we find that
\be
\Delta t_{\rm e}=5t_{\rm e}[a\sin(\omega t)+\frac{\Delta
T_{\rm e}}{2T_{\rm e}}].
\label{} 
\ee
Since the adjustment time of the electron temperature is short, the heating rate of the 
electrons must equal the (Compton) cooling rate $\dot q_{\rm cool}$ at any instant:
\be
\dot q_{\rm heat}=\frac{nk_{\rm B}T_{\rm e}V}{t_{\rm e}}=\dot q_{\rm
cool}.\label{bal}
\ee

\subsection{Evaluation of the work integral}

The oscillations involve the pressure force and gravity. In an ion supported flow, both these
forces are concentrated in the protons. Hence the cooling/heating rate $\dot q$ in the work
integral involves only the proton temperature and the distance from the central mass
\be \dot q =\dot q(T_{\rm p}, r(t)). \label{dotq}\ee 
The proton temperature $T_{\rm p}$ of a mass element at distance $r(t)$ changes with time due
to three factors, the compression/expansion of the gas, energy transfer to the electrons, and the
viscous heating. Variations in viscous heating as a possible excitation  mechanism of oscillations in
accretion disks has been studied before (Kato 1978; Papaloizou \& Stanley 1986; Kley et al.
1993). The maximum growth rate due to this process is of order $\alpha\Omega$, where
$\alpha$ is the disk viscosity parameter. Assuming $\alpha$ to be small,  perhaps of order 0.01-0.1 as
indicated by numerical simulations of magnetic accretion disk turbulence (e.g. Brandenburg et al. 1997; 
Armitage 1998), this contribution is small
compared with the growth rate we find below due to the other interaction processes, hence we
will neglect variations in viscous dissipation during the oscillation cycle in the following (see also
Sect. \ref{combi}).

Assuming we are looking at a slowly damped or growing mode, the temperature variation can
be written as $\Delta T=\Delta T_{\rm ad}+\epsilon \Delta T_{\rm cool}$, where $ \Delta T_{\rm
ad}(t)$ is the adiabatic temperature variation, $\Delta T_{\rm cool}$ is the temperature change
due to the variation in energy transferred to the electrons, and $\epsilon$ a small number.  
Expanding in $\epsilon$, the work integral can be evaluated by setting $\Delta T=\Delta T_{\rm
ad}$ in  Eq. (\ref{dotq}),  that is, it can be found to lowest order using only the adiabatic 
variation of the proton temperature.

 In practice we evaluate the work integral in a highly discretized way:  the integral is
represented by the sum of two values: the integrands at the phases of maximum contraction and
maximum expansion. For a linear (sinusoidal) oscillation, evaluation at a single known phase in the
oscillation would be enough to determine its amplitude, and would be sufficient to determine the
work integral. By evaluation at the phases of both maximum and minimum expansion, nonlinearity
due to asymmetry between the expanded and contracted phases can be taken into account to
lowest order. This is somewhat important for the present problem because an asymmetry is in
fact expected since, for large oscillatory amplitude, the ISAF in its contracted phase can 
actually detach from the
cool disk. For the two phases, the cooling rate is evaluated by an iterative
procedure, varying the electron temperature until the balance  Eq. (\ref{bal}) is satisfied.

\subsection{The Monte-Carlo code}

In order to study both the excitation and the damping mechanisms that take place in the ISAF 
quantitatively, the Comptonization of soft radiation has to be calculated in detail. For this 
we have performed Monte Carlo simulations following Pozdnyakov et al. (1983). For specific values 
of the electron temperature $T_{\rm e}$, the optical depth $\tau$, location and energy of the soft 
photons, we calculate the cooling rate of the ISAF as well as the X-ray spectrum produced.

 The location of the soft photons depends on the mechanism that produces them. In the case of
thermal synchrotron photons, they are expected to be emitted in the region close to the
black hole, with an energy distribution that strongly peaks at a few eV. For simplicity we
assume that all synchrotron photons come from the center of the ISAF. If the soft photon flux
is a result of disk irradiation caused by hard X-rays coming from the inner hot flow, then the
location of their emission and their energy is calculated by the iterative method that is 
described in section 5.1. Finally in the `proton illumination' case, the soft flux comes from
the inner part of the thin disk where the disk interacts with the ISAF (at radii $R_{\rm tr}<r<
R_{\rm out}$). The photon energy in this case is given by Eq. (\ref{Tdisc}).     

The Monte Carlo method follows the path of a large number of photons and makes a statistical 
description of the radiation transfer inside the scattering medium. The calculation takes properly 
into account the quantum mechanical and special relativistic effects of the microphysics of the scattering. 
We make sure that the statistical sample (i.e. the number of photons) is 
large enough to keep statistical noise at a low level.
   
\section{Results}

Whether small amplitude radial oscillations of the ISAF, once appearing, will increase in amplitude 
or die out depends on the origin of the soft photons. As discussed in section \ref{phsrc}, there are 
three possible sources of soft photons: the reprocessing of hard (inverse Compton) photons intercepted 
by the cool disk (the `X-ray irradiation' model), photons produced by ion-heating of the cool disk 
(`proton illumination'), and internal (synchrotron) photons generated in the ISAF. The conditions for 
excitation are different in each case. Although all these processes may contribute, we study their 
effects separately. We start with the irradiation model.

\subsection{X-ray irradiated cool disk} 

Parameters of the ISAF are its equilibrium values of electron temperature and optical depth, using 
Fig.\ 1 as a guide. The degree of overlap of the ISAF over the cool disk in the equilibrium 
state, $f=R_{\rm out}/R_{\rm tr}-1$, is a third parameter of the problem.
We make a guess of the initial radial temperature profile of the disk surface and simulate the emission of cool 
photons from the disk and the upscattering of these soft photons into the ISAF with a Monte Carlo calculation. 
Part of the hard X-rays produced by inverse Compton scattering in the ISAF will meet the thin disk in their 
scattering random walk and get either absorbed or reflected. The absorbed X-ray flux, assumed to be 
thermalized into a black body, determines the new temperature profile of the disk surface. The simulation is 
repeated using the new temperature profile several times until it converges (within numerical 
uncertainties imposed by the statistical nature of the Monte Carlo method). The cooling rate of the 
ISAF in the final run is recorded.

The next step is to expand the medium by a small amount. The optical depth of the hot medium is reduced 
 in accordance to Eq.\  (\ref{dtau}) for $\sin(\omega t)=1$ (maximum expansion), while the temperature 
of the hot medium on the expanded phase is given a trial value. Following the procedure described in the 
previous paragraph, the temperature profile of the disk and the cooling rate of the ISAF $\dot 
q_{\rm exp}$ are computed. Then we iterate varying the temperature of the ISAF until Eq.\ (\ref{bal}) 
is satisfied. The same procedure can be applied to maximum contraction of the ISAF  to compute its cooling 
rate $\dot q_{\rm contr}$.

There are two simplifications made during this computation. First, we have ignored the reflected component. 
Detailed spectral predictions are out of the scope of this work and, on the other hand, the reflected spectrum 
lies in the X-ray regime and is not expected to contribute much in the ISAF cooling (in which we are mainly 
interested). Second, we kept the fraction of the flux that the disk reflects in the different phases of 
the oscillation constant. This, however, may not be the case (see for example Nayakshin et al. 2000).

Keeping this in mind, we find that in the irradiation model $\dot q_{\rm exp}-\dot q_{\rm 
contr}<0 $ for the entire range of the parameter space $\tau,{}T_{\rm e},{} f$ that we have 
investigated. It seems to be a robust result that \emph{when soft photons are product of 
reprocessing of X-rays by the surface of the thin disk, small amplitude radial oscillations of the ISAF 
are damped}.

\subsection{Proton illumination}

In the proton illumination model the soft photon flux comes from the overlap region only. Part of the 
proton illumination energy (about half, like in the Haardt-Maraschi energy balance) is reprocessed into 
blackbody radiation and the rest is 
emitted in the X-ray band due to the `warm' layer that forms in the overlap region (see Deufel \& Spruit 
2000 for details). Here we will ignore the X-rays coming from the warm layer. The total soft luminosity is
\be 
L_{\rm soft}\simeq \frac{1} {2} F_{P} A,
\label{lsoft}
\ee
where $F_{P}$ is given by Eq. (\ref{flux})  and $A=2\pi (R_{\rm o}^2-R_{\rm tr}^2)$ is the overlap surface 
area (the factor 2 accounts for the two sides of the disk). Taking the Lagrangian derivative of Eq.
(\ref{lsoft}), we arrive at the result
\be
\Delta L_{\rm soft}= 2 L_{\rm soft} 
\bigg( \frac{3K^2-2}{1-K^2}\bigg) a \sin(\omega t), 
\label{}
\ee 
where $K=R_{\rm tr}/R_{\rm out}$, or $f=1-K$. From this equation it follows that for $K^2<2/3$ (which 
corresponds to $f>0.18$), the soft flux decreases on expansion and increases on contraction of the ISAF. 
So for large overlap fraction, $f>18 \%$, any small amplitude radial oscillations are expected to be 
damped (see also the discussion of the the synchrotron soft photon case  in Sect. 3.1). Of more interest 
is the regime of $f=1 -R_{\rm tr}/R_{\rm out}<10 \%$.

From calculations with different fractional overlap $f$, we find that \emph{there is a characteristic 
value $f_{\rm o}$ below which oscillations are excited}. This is illustrated in Fig. 2. It shows, 
as a measure of the work integral, the quantity $ Y=(\dot q_{\rm exp}/\dot q_{\rm contr}-1)/a$, the 
fractional variation of the cooling rate, divided by the relative amplitude $a$ of the oscillation. 
It is proportional to the growth rate $\eta$ of the oscillations (see appendix A).  For the example 
shown ($\tau=1.5$ and $T_{\rm e}=130$ keV) the regime of excitation is $f\le 0.07$. The results remain 
qualitatively the same for different choices of $T_{\rm e}$ and $\tau$.

Until now we have only dealt with small amplitude oscillations. What limits the growth of these 
oscillations? To answer this questions we have also calculated the variation of the cooling rate 
for larger amplitudes. The value of the amplitude at which the work integral changes sign is a 
good estimate of the amplitude reached by the oscillation. Again we approximate the work integral 
by the sum of the integrand in the expanded and contracted phases.

\begin{figure}
\resizebox{\hsize}{!}{\includegraphics[]{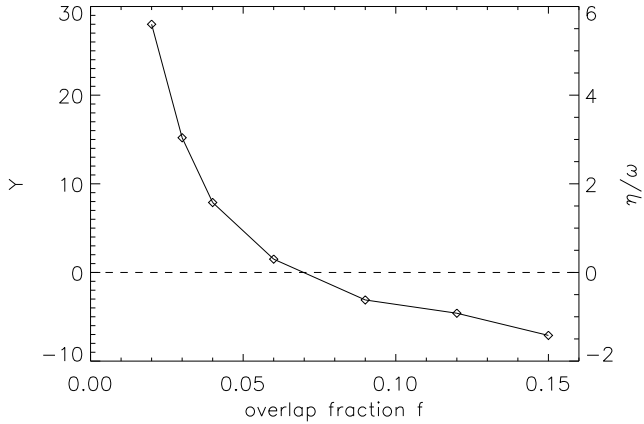}}
\caption[]{The growth/damping rates of small amplitude oscillations
for different overlap fractions $f=(R_{\rm out}-R_{\rm
tr})/R_{\rm out}$ in the proton illumination case. The left vertical axis shows the work 
integral $ Y=(\dot q_{\rm exp}-\dot q_{\rm contr})/(a\dot q_{\rm eq})$, the right axis the 
corresponding growth rate $\eta$ in units of the frequency $\omega$ of the oscillation. 
For $f<0.07$ the work integral is positive and the oscillations are excited. 

}  
\label{fig2}
\end{figure}
\begin{figure}
\resizebox{\hsize}{!}{\includegraphics[]{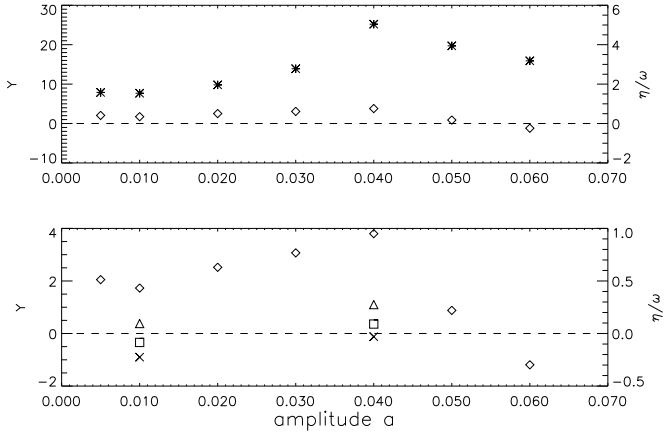}}
\caption[]
{Work integral  $Y$ of large amplitude oscillations. Upper panel: the proton illumination case 
(asterisks) and the case where there is also a soft synchrotron flux with $L_{\rm illum}/L_{\rm syn}=230$ 
(diamonds). Lower panel:  $L_{\rm illum}/L_{\rm syn}=$ 230 (diamonds), 163 (triangles), 150 (squares), 
138 (X). Note that for  $L_{\rm illum}/L_{\rm syn}=150$, small amplitude oscillations are damped, 
while larger amplitude oscillations ($a\sim 4\%$) are excited.}
\label{fig3}
\end{figure}

Fig.\ \ref{fig3} (upper plot) shows the results for an equilibrium overlap fraction $f=0.04$. 
The growth rate has a maximum for $a =0.04$. 
When $a >0.04$ the ISAF, during part of the contracting phase, does not interact with the disk 
($f=0$) and X-ray flux ceases. Note that the work integral is positive for all amplitudes: there is no limit to 
the growth in this simplified case. More realistically, hydrodynamic nonlinearities ignored here 
would limit the oscillation. Also, other sources of soft photons will have a limiting effect, as 
discussed  below.

\begin{figure}
\includegraphics[width=0.35\textwidth,height=0.4\textheight,angle=270]{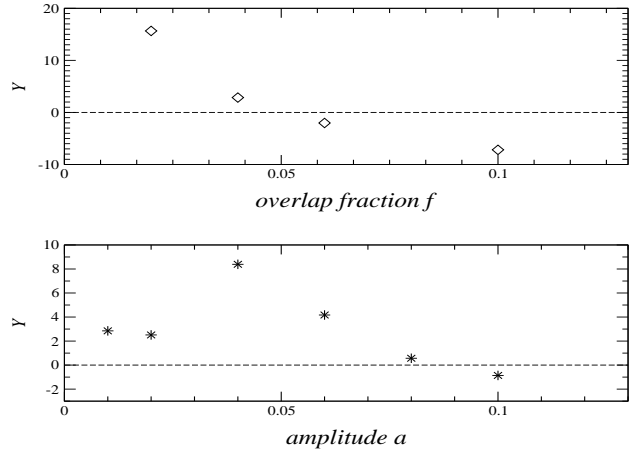}
\caption[]
{Work integral for the case where both proton illumination and irradiation are taken into account.
Upper panel: small amplitude oscillations as a function of overlap fraction $f$. For
$f\simless 0.05$ small amplitude oscillations are excited. Lower panel: work integral as a function 
of oscillation amplitude, 
for fixed overlap fraction $f=0.04$. In this case the oscillation grows up to $a\sim 0.09$.
These results are for ISAF temperature $k_{\rm B}T_{\rm e}=150$ keV, and $\tau=1$. 
\label{figex}}
\end{figure}

\subsection{Combination of different soft photon sources}

 \label{combi}
The situation described in the previous section leaves out some physics. Synchrotron photons 
from the ISAF and soft photons produced by irradiation of the cool disk are expected to also 
play a role in determining the amplitude of the oscillation, since we have already shown that 
they act as damping mechanisms. As a first step in this direction, we now include synchrotron 
photons into the simulation. The amount of synchrotron soft photons produced by the ISAF depends 
on the unknown field strength. Hence we treat the synchrotron luminosity $L_{\rm syn}$ as a free 
parameter. The soft flux due to proton illumination in this example is a black body with 
$T_{\rm illum}=100$ eV, while the typical  energy of the synchrotron photons is set at $\sim 3$ 
eV. The values of the other parameters are $T_{\rm e}=130$ keV, $\tau=1.5$, $f=0.04$.

For $L_{\rm illum}/L_{\rm syn}=230$ (Fig. \ref{fig3}), the oscillation amplitude grows up to  
$a \sim 0.05$ and stabilizes at that level. A weak synchrotron flux can limit the 
amplitude's growth to values slightly larger than the overlap fraction. However, if the synchrotron
flux is strong enough, $L_{\rm illum}/L_{\rm syn}\simless 140$, the oscillation does not get 
excited at all.

An interesting feature of the combined synchro\-tron-proton illumination model for the soft photons
 is the appearance, in this example, of `hard' oscillations. As Fig. \ref{fig3} shows, small 
amplitude oscillations are damped, while the work integral is positive for oscillations with larger 
amplitudes, $a \sim 0.04$, if the synchrotron flux is in the range 
$140 \simless L_{\rm illum}/L_{\rm syn} \simless 155$.

In this work we have ignored any variation of the viscous heating during the oscillation. However, 
viscous heating can trigger/damp oscillations on the viscous time scale (e.g. Kley et al. 1993). 
Following the $\alpha $ viscosity prescription (Shakura \& Sunyaev 1973) and taking that $H\simeq 
R$ for an ISAF, we have $t_{\rm visc}\sim t_{\rm dyn}/\alpha $, where $H$ is the 
thickness of the ISAF and $t_{\rm dyn}=1/\omega$ is the dynamical time scale. So the viscous 
excitation/damping rate is ${\eta}_{\rm visc}/\omega \sim \alpha$. Comparing this with the rates 
due to radiative cooling (Fig. 2, 3), we see that for $\alpha \simless 0.1$ the effects of viscous 
heating on excitation can be safely neglected.

\subsubsection{Reprocessed hard X-rays}

 We now consider the case when the soft photon flux is produced by the joint effects of
proton illumination and reprocessing of the hard X-rays from the ISAF. Keeping the 
proton illumination component of the soft flux as a black
body with $T_{\rm illum}=100$ eV and, calculating the temperature profile of the disk due to
X-ray irradiation with the method described in  Sect. 5.1, we study whether small amplitude
oscillations are excited or damped for different overlap fractions $f$. In this example we have
fixed $k_{B}T_{\rm e}=150$ keV and $\tau =1$.

The results are presented in Fig.\ \ref{figex}, showing the growth rate as a function of the 
overlap fraction and oscillation amplitude. For $f\simless 0.05$ small amplitude oscillations are 
excited. Fixing the equilibrium overlap fraction $f$ to 0.04, we have also calculated the 
variation of the cooling rate of the ISAF for larger oscillation amplitudes (lower panel). 
The maximum growth rate occurs at an amplitude $a=0.04$. The amplitude of the oscillation 
can be estimated from the change of sign of the grwoth rate as a function of amplitude, yielding 
$\alpha_{\rm max}\sim 0.09$.

\subsubsection{Viscous dissipation in the cold disk as a source of soft photons}

A source of soft photons ignored so far is viscous dissipation in the thin disk. 
Including this soft-photon source, using standard thin disk expressions (Shakura \& Sunyaev 1973),
we find that the results presented in the previous sections remain essentially the same. This is so
because the intrinsic luminosity of the disk has only weak triggering/damping effects on the oscillations
of the ISAF.

\subsection{Spectral evolution during the oscillation} 
\label{result} 
From the discussion in the previous sections, it is evident that if the overlap fraction between the thin 
disk and the ISAF is small enough (typically smaller than ~7\%), radial oscillations of the ISAF can be 
excited. 
The overlap region is a source of hard X-rays (Deufel et al. 2001, 2002). The ISAF itself, however, 
may actually be the main X-ray producing region, and its oscillations are 
expected to have direct implications on the shape of the emitted spectrum.

On expansion of the ISAF the optical depth, the electron temperature and hence the Compton 
$y$ parameter drop. The soft photon input on the other hand increases. As a result the spectrum becomes 
softer while the total X-ray flux increases.

\begin{figure}
\resizebox{\hsize}{!}{\includegraphics[]{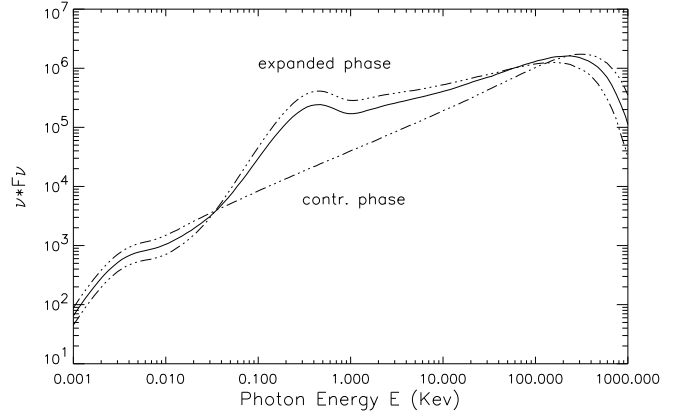}}
\caption[]
{Spectra predicted by the combined model of Fig. 3 for different phases of the oscillation. 
Solid line corresponds to the equilibrium position while dash-dotted lines correspond to maximum expansion 
and maximum contraction. $L_{\rm illum}/L_{\rm syn}=230$ (see diamonds in Fig. 3). 
\label{fig4}
}
\end{figure}

The opposite happens in the contracting phase when the source becomes harder and the 
X-ray flux decreases. So the X-ray spectrum exhibits a characteristic pivoting behavior which is 
shown in Fig.\ \ref{fig4}. In this figure (which corresponds to the case where both proton 
illumination and synchrotron photons are taken into account) we have plotted the spectrum computed 
by the Monte Carlo code in equilibrium position, in maximum expansion and in maximum contraction.

One clear conclusion from Fig.\ \ref{fig4} is that the oscillations
are predicted to be soft i.e. to have more power in soft X-rays
than in hard X-rays. A second conclusion is that a simultaneous oscillation (but
of opposite phase) should be observed at optical/infrared wavelengths,
if the synchrotron soft flux accounts for a large fraction of the
total  optical/infrared luminosity of the source. Implications from
these conclusions are discussed in the next sections.           

\section{Consequences/Predictions}

Radial oscillations of the ISAF excited by the interaction between an ISAF and a cool disk have 
a number of characteristics from which they can be identified and conclusions can be 
drawn about the accretion geometry. The most promising source of excitation turned out to be the 
interaction between ISAF and cool disk via proton illumination. For this mode of interaction, 
the model predicts the following.

i) The QPO frequency is predicted to be close to the Kepler frequency in the outer edge of 
the ISAF. For frequencies characteristic of the oscillations (around 1 Hz), the ISAF must 
extend to $R_{\rm out}\sim 100 R_{\rm s}$. This is comparable to the extent of the ISAF in 
the low/hard state in Cygnus X-1, as deduced by Gilfanov et al. (2000) from Fourier 
frequency-resolved spectroscopy.

ii) As shown in Sects. (5.2), (5.3) the excitation of the oscillations can be very fast, 
with $\eta /\omega\sim 1$. This implies that they are expected to show non-sinusoidal 
behavior and higher harmonics. Under these conditions our assumptions about almost sinusoidal 
oscillations are not valid any more. The main conclusions, however, are expected to stay 
qualitatively the same.

iii) There are cases with `hard' or finite amplitude instability, in which small amplitudes 
are damped but finite amplitudes can grow rapidly.

iv) If synchrotron flux accounts for part of the soft flux that is Comptonized, then infrared 
and/or visible modulations of the same frequency with the QPOs are predicted. The modulations 
will more likely be observed if synchrotron flux accounts for a large fraction of the total 
source emission in the relevant wavelengths.

v) Excitation of the oscillations occurs for overlap fraction of the ISAF and the cool disk less 
than about ~7\%{}. If the observed QPOs are excited by the processes described here, there must 
be a rather sharp transition from the outer thin disk to the inner hot flow.

A final conclusion is that the QPOs are predicted to be soft, showing larger rms amplitudes in the 
soft X-ray bands than in the hard ones. This, at first sight, seems to conflict with the usual 
observational trend that QPOs are hard. There are, however, several observations of sources in the 
low/hard state where soft QPOs have been observed. During the decay after the 2000 outburst of XTE 
J1550--564 (Kalemci et al. 2001), or in the hard state of Cygnus X-1 (Rutledge et al. 1999) soft 
oscillations have been observed with frequencies around ~1 Hz. This suggests that there are several 
mechanisms contributing to QPO excitation, but makes clear that other mechanisms than those described 
here need to be explored.
     
\section{Summary and discussion} 

The implications of any model that attempts to describe QPOs depend strongly on the assumed accretion 
geometry. In this work we have assumed that the accretion flow consists of a standard thin disk which 
is truncated in a radius $R_{\rm tr}$ and a hot two-temperature ion supported flow (ISAF) inside this 
radius. Due to its tendency to spread outward viscously, the ISAF extends to a radius 
$R_{\rm out}>R_{\rm tr}$, resulting in partial overlap of the two flows.

We have focused on the frequency and the excitation/damping mechanisms of the fundamental mode of radial 
oscillations of the ISAF. The oscillatory frequency of this mode is close to the Kepler frequency in 
the outer edge of the ISAF. We find that the main contributor to the excitation of these oscillations is 
the variation of cooling rate of the ISAF in the different phases of the oscillation. When the ISAF cools 
more efficiently in the expanded phase than the contracted phase, an oscillation is excited.

We have assumed Compton cooling to be the dominant cooling physical process
of the ISAF. Since the  source of the Comptonized `seed' photons depends on 
the still unknown details of the accretion geometry, we have considered three 
different physical mechanisms that have previously been
proposed to account for the soft photon flux: i) synchrotron flux due to frozen in 
magnetic fields in the ISAF, ii) irradiation of the thin disk by X-rays from the
ISAF (Haardt \& Maraschi 1993) and iii) proton illumination of the disk in the
overlap region (Deufel \& Spruit 2000). 

We have developed a Monte Carlo code to calculate the Compton scattering and
to calculate the cooling of the ISAF, as well as the emergent X-ray spectra during
the oscillations. We find that soft photons produced by synchrotron and by reprocessing 
of the hard flux act as damping 
mechanisms of the oscillations. On the other hand, proton illumination excites oscillations if the overlap 
fraction $f$ between the disk and the ISAF is small enough (of the order of ~7\%). Also models that 
combine proton illumination with synchrotron or irradiation soft flux can produce (if some constraints 
are fulfilled) radial oscillations with amplitudes of the order of the overlap fraction, i.e. of the 
order 10\%.

The oscillations are predicted to produce a QPO signature in the X-ray power spectrum that has more 
power in the soft X-rays than in the hard X-rays (soft QPOs). This leads us to identify the low 
frequency QPOs observed in the decay of the 2000 outburst of XTE J1550--564 (Kalemci et al. 2001) 
and in the hard state of Cygnus X-1 (Rutledge et al. 1999) within the framework of our model.

\begin{acknowledgements}
 We would like to thank the
anonymous referee for useful suggestions that greatly improved the manuscript.
 Giannios acknowledges partial support from the EC Marie Curie Fellowship HPMTCT 2000-00132
 and the Program `Heraklitos' of the Ministry of Education of Greece. 
\end{acknowledgements}

\appendix
\section{The decay/excitation time scale.}

The damping/growth rate $\eta$ of the amplitude of an oscillation can be
defined as
\be
\eta =1/2\frac{\langle\dot W\rangle}{\langle E_{\rm osc} \rangle},
\label{eta}
\ee
where $\langle\rangle$ expresses the average value of a quantity over a period, $E_{\rm osc}$ is the oscillatory energy and $\dot W$ is the rate of energy gain of the oscillation. If $\langle\dot W\rangle$ is a negative number, then $\eta <0$ and corresponds to the damping rate of the oscillatory amplitude, while if $\langle\dot W\rangle$ is positive, then $\eta >0$ and corresponds to the excitation rate.

The total energy of the oscillation equals the kinetic energy of the
oscillation when the medium crosses the equilibrium position. With
Eq. (\ref{wint}), integrating over the volume of the ISAF we find
\be 
E_{\rm osc}=\frac{2\pi \tau m_{\rm p} {a}^2{\omega}^2 R_{o}^4}{5 \sigma_{
\rm T}}.
\label{}
\ee

For nearly sinusoidal oscillations, $\langle\dot W\rangle$ is given by the integral
\be
\langle\dot W\rangle=-\frac{1}{T} \int_m dm \int_0^T
(\gamma-1)\frac{\Delta\rho}{\rho}\Delta \dot q dt,
\label{}
\ee
where $\Delta\dot q$ is the variation of the cooling rate per unit
mass of the ISAF. In general $\Delta\dot q$ is function of the space
coordinates and time. Taking the cooling rate constant over the volume
of the medium (numerical checks show that this is rather
realistic) and to vary sinusoidally with time, we have
\be
\Delta\dot q=\dot q_{\rm amp} \sin(\omega t+\phi).
\label{ddq}
\ee
With $\dot q_{\rm amp}=(\dot q_{\rm exp}- \dot q_{\rm
contr})/(2M_{\rm ISAF})$ being
the amplitude of the oscillation of $\dot q$. Also $\dot q_{\rm exp}$ / $\dot
q_{\rm contr}$ is the cooling rate of the ISAF in maximum
expansion/contraction and $M_{\rm ISAF}$ is the mass of the ISAF.

 Expression (\ref{ddq}) contains a phase $\phi$. For $-\pi /2<
\phi< \pi /2$, the oscillation is excited, while for $\pi/2<\phi <\
3/2\pi$, the oscillation is damped. We will focus on the extreme
cases $\phi=0$ and $\phi=\pi$. (The first corresponds to
$\dot q_{\rm amp}>0$ and the second to $\dot q_{\rm amp}<0$).

Integrating (\ref{ddq}) over a period and over the mass of the ISAF we find
\be
\langle \dot W\rangle = 3(\gamma-1)a\frac{\dot q_{\rm exp}- \dot
q_{\rm contr}}{4}.
\label{}
\ee
So, from (\ref{eta}) we can estimate the damping/growth rate of the
amplitude of the oscillation. Of more interest is the quantity $\eta
/\omega$, where $\omega$ is the the frequency of the oscillation, taken
to be the Keplerian frequency at the outer edge of the ISAF.
    
This yields
\be
\frac{\eta}{\omega}=\frac{15}{4}\frac{(\gamma-1)}{\tau}\frac{c}{(GM/R_{\rm
out})^{1/2}}
\frac {L_{\rm x}}{L_{\rm Edd}} \frac{(\dot q_{\rm exp}- \dot q_{\rm contr})/\dot
q_{\rm eq}}{a}
\label{}
\ee
where $L_{\rm x}$ is the X-ray luminosity of the source and $\dot q_{\rm eq}$
is the the cooling rate of the ISAF in its equilibrium position.
In order to derive this equation we have also used the expression for
the Eddington luminosity $L_{\rm E}=4\pi GMm_{\rm p} c/{\sigma}_T$ and
the fact that $L_{\rm x} \simeq \dot q_{\rm eq}$ (since inverse Compton is 
responsible for both the cooling of the ISAF and the X-ray luminosity
of the source).    

The last expression can also be written more conveniently as
\be
\frac{\eta}{\omega}=\frac{3.5}{\tau}r_{\rm out}^{1/2}\frac{L_{\rm x}}{L_{\rm
E}}\frac{(\dot q_{\rm exp}- \dot q_{\rm contr})}{a\dot q_{\rm eq}}.
\label{}
\ee
Note that the growth rate of the oscillation is proportional to{} $Y
= (\dot q_{\rm exp}-\dot q_{\rm contr})/(a\dot q_{\rm eq})$.

\end{document}